\documentclass[12pt]{JHEP3}
\usepackage{graphicx}
\usepackage{epsfig}
\usepackage{amssymb}
\usepackage{bm}
\usepackage{amsmath}
\newcommand{\bea}{\begin{eqnarray}}
\newcommand{\eea}{\end{eqnarray}}

\newcommand{\bra}{\langle}
\newcommand{\ket}{\rangle}

\newcommand{\nn}{\nonumber}

\newcommand{\bfg}{{\bf \Gamma}}
\newcommand{\mca}{{\mathcal A}}
\newcommand{\mcf}{{\mathcal F}}
\newcommand{\rmd}{{\mathrm d}}

\makeatletter
\@addtoreset{equation}{section}
\makeatother

\title{
Ramond-Ramond couplings of D-branes
}
\author{
Koji Hashimoto$^{1,2,*}$, 
Sotaro Sugishita$^{3,\#}$ and Seiji Terashima$^{4,\dagger}$
\\

$^1$
{\it Department of Physics, Osaka University,
Toyonaka, Osaka 560-0043, Japan}\\
$^2$
{\it Mathematical Physics Lab., RIKEN Nishina Center,
Saitama 351-0198, Japan}\\
$^3$
{\it Department of Physics, Kyoto University, Kyoto 606-8502, Japan}\\
$^4$
{\it Yukawa Institute for Theoretical Physics, Kyoto University, Kyoto 606-8502, Japan}\\
E-mail: $^*$ \email{koji(at)phys.sci.osaka-u.ac.jp}\\ 
E-mail: $^\#$ \email{sotaro(at)gauge.scphys.kyoto-u.ac.jp}\\
E-mail: $^\dagger$ \email{terasima(at)yukawa.kyoto-u.ac.jp}\\ 
}

\abstract{Applying supersymmetric localization for superstring worldsheet
theory with  ${\cal N}=(1,1)$ supersymmetries on a cylinder 
and with arbitrary boundary 
interactions, we
find the most general formula for the Ramond-Ramond (RR) coupling of D-branes.
We allow all massive excitations of open superstrings, and find that only a finite number of them
can contribute to the formula. The formula is written by 
Quillen's 
superconnection which includes higher form gauge fields, 
and the resultant general Chern-Simons terms
are consistent with RR charge quantization. 
Applying the formula to boundary string field theory
of a BPS D9-brane or a D9-antiD9 brane system,
we find that any D9-brane creation via massive mode condensation is impossible. 
}

\preprint{
{\normalsize KUNS-2532} \\
{\normalsize OU-HET-845} \\
{\normalsize RIKEN-MP-103}\\
{\normalsize YITP-14-107}\\
}


\begin{document}
\setcounter{page}{1}

%%%%%%%%%%%%%%%%%%%%%%%%%%%%%%%%%%%%%%%%%%%%%%
\section{Introduction: Most general Ramond-Ramond charge formula}

Identifying the 
Ramond-Ramond (RR) coupling of D-branes in superstring theory was the ignition of 
the second revolution by J.~Polchinski \cite{Polchinski:1995mt}. 
Since then, tremendous amount of research has been carried out based on the D-brane RR-charge formula: the coupling between the RR field in the bulk and the massless open string degrees of freedom on 
multiple D$p$-branes is \cite{Li:1995pq, Douglas:1995bn}
\begin{eqnarray}
S_{\rm RR}=T_q \sum_p \int  \; C_{\rm RR}^{(p+1)}\wedge {\rm Tr} \; e^{2 \pi \alpha' F}
\label{RR}
\end{eqnarray}
where $F=dA-i A^2$ is a two-form field strength of a massless Yang-Mills gauge 
field on the D-branes. 
$T_q$ is the tension of a D$q$-brane and
$C^{(p+1)}_{\rm RR}$ is the RR $(p+1)$-form field, where $p$ is an odd (even) integer for type IIB (IIA) superstring theory. Since the RR coupling formula (\ref{RR})
generates Chern characters for the Yang-Mills fields, the formula is consistent with
RR charge quantization. 

The full structure of the RR charge formula has not been addressed, because of
complexity of the string excitations. In fact, the formula (\ref{RR}) should be able to be
generalized to include all the open string excitations, even if we restrict our attention
to the lowest tree-level in perturbation of string theory (that is a disk amplitude). 
In this paper, we provide a completely general formula for the RR coupling of D-branes, with arbitrary number of all massive open string excitations.

A part of the generalized RR coupling was widely known 
\cite{Kraus:2000nj,Takayanagi:2000rz}
in the context of tachyon condensation in string theory. Starting with a pair of a 
D-brane and an anti-D-brane (or a non-BPS D-brane), 
one can include tachyon modes in addition to the
massless modes. The RR coupling formula including the tachyon modes was given 
in the context of boundary string field theory (BSFT) \cite{BSFT}. 
For superstring theory,
the action of the off-shell open string field theory is given by just a disk partition function with arbitrary boundary vertex insertions allowed \cite{KMM2,ZS,Andreev-Tseytlin}. 
The resultant RR coupling formula is written with the Yang-Mills fields and the tachyon fields, and it was used to show Sen's conjecture \cite{Senconje1,Senconje2,Senconje3,Senconje4,Senconje5} 
on D-brane annihilation via tachyon condensation.\footnote{
In the BSFT, the decent relations in the Sen's conjecture 
were also proved \cite{AST}.}

Any non-perturbative formulation of superstring theory needs to be capable of
describing a D-brane creation, as well as the D-brane annihilation. As one can naively guess, to describe a D-brane creation, it is plausible to have a condensation of 
open string massive modes, rather than the tachyon condensation. The mass squared
of the open string modes signals the direction of the energy uplift / decrease.
So, to investigate 
whether open superstring theory can describe multiple D-brane creation in
its off-shell configuration space or not, we need to generalize the previously known 
RR charge formula (\ref{RR}) and an analogous one including the tachyons, to
the one with all massive open string modes.\footnote{A treatment of all massive open string modes in BSFT was studied in detail in \cite{Hashimoto:2004qp}. 
For a bosonic BSFT, condensation of massive modes was analyzed in 
\cite{Hashimoto:2012ig}.}

The necessary technique to calculate the most general RR charge formula
is indeed ready, thanks to the recent developments on supersymmetric
localization technique. (See, for example, \cite{Pestun}-\cite{Terashima:2014yua}.
 For manifolds with boundaries, see \cite{Sugishita:2013jca}-\cite{Yoshida:2014ssa}.) 
As we mentioned, the RR charge formula at off-shell superstring theory
is nothing but a worldsheet partition function. We apply the supersymmetric localization 
to the supersymmetric worldsheet theory with arbitrary number of arbitrary massive
open string interactions at the worldsheet boundary.

The type II superstring worldsheet theory has ${\cal N}=(1,1)$ supersymmetries.
We are interested in a flat target space-time for the moment, as we need to know
what is the RR field configuration; at the flat spacetime, the RR field 
can be constant, which suits our purpose. Therefore the 
worldsheet theory has just chiral multiplets in 1+1 dimensions.
Nevertheless, we need to allow arbitrary interactions at the worldsheet boundary,
which breaks only a half of the ${\cal N}=(1,1)$ supersymmetries on the worldsheet. 
We develop the localization technique for the theory and obtain an exact partition function
with the arbitrary boundary  interactions.\footnote{
Note that these generically break the space-time SUSY.}
Resultantly, we obtain {\it the most general RR charge formula} :
\begin{eqnarray}
S_{\rm RR}=T_q \sum_p \int  \; C_{\rm RR}^{(p+1)}\wedge {\rm Str} \; e^{2 \pi \alpha'  {\cal F}}
\label{genRR}
\end{eqnarray}
This most general RR charge formula is written beautifully 
in terms of Quillen's superconnection ${\cal A}$ \cite{Quillen,Berlie}, with 
the field strength ${\cal F}\equiv d {\cal A}- i{\cal A}^2$. 
The superconnection includes higher form fields which are massive open superstring excitations. It also includes tachyon part 
for the case of the D9-antiD9 system
which was conjectured in \cite{Kennedy:1999nn} and derived in 
\cite{Kraus:2000nj,Takayanagi:2000rz}.

The open superstring massive modes show 
up in our general RR charge formula such that the quantization of the RR charge 
is ensured. This is natural but
surprising, as only a limited number of open string excitation modes can enter
the RR charge formula. In fact, only a finite number 
of all the massive modes can appear 
in the general formula. For the RR charge formula for 
a BPS D-brane, only five of them (including the massless modes) show up.

From the exact partition function of the string worldsheet as a boundary superstring field theory, 
we can provide a proof of no D9-brane creation, starting from a D9-brane, 
or a D9-antiD9 pair.
So, even with massive field condensation, one never obtains a creation of the D9-brane in the context of BSFT.

The organization of this paper is as follows. In section \ref{sec2}, we provide 
a supersymmetric localization for the ${\cal N}=(1,1)$-supersymmetric worldsheet theory with chiral matter multiplets. Then in section \ref{sec3}, we study the generic 
boundary interaction of the worldsheet with the localization, and find that it is impossible to generate a D9-brane charge starting from a single BPS D9-brane 
or from a D9-antiD9 pair. In section \ref{sec4}, we provide the most general 
RR charge formula with arbitrary massive open string excitations, and describe its relevance to Quillen's superconnection.

%%%%%%%%%%%%%%%%%%%%%%%%%%%%%%%%%%%%%%%%%%%%%%

\section{Localization in $(1,1)$-supersymmetric worldsheet theory}
\label{sec2}

The localization computation in Euclidean 2-dimension space has been considered
for ${\cal N}=(2,2)$ SUSY theories.
Here, we will consider an ${\cal N}=(1,1)$ SUSY theory on a 2-dimensional flat 
cylinder with a finite length.\footnote{Superstring partition functions at tree level are considered
with a disk worldsheet. In this paper we treat cylinder instead of the
disk as in \cite{AST},
because the SUSY boundary conditions are simple.
With appropriate boundary conditions, the disk and cylinder partition
functions gives a same result for the on-shell space-time fields
which correspond to conformal boundary interactions.
For the off-shell fields, we need a field redefinition.
}
In this setting, we can include {\it all fluctuations} on D-branes,
which are not needed to be space-time supersymmetric.
We will take the notations and conventions which are 
used in the Polchinski's text book
and we will take $\alpha'=2$ below.
In the 2d bulk which is assumed to be flat, 
we have a following bosonic superfield:
\begin{eqnarray}
 {\bf X}^\mu =X^\mu+ i \theta \psi^\mu + i \bar{\theta} \tilde{\psi}^\mu
+ \theta \bar{\theta} F^\mu,
\end{eqnarray}
where $\theta$ and $\bar{\theta}$ are independent 1-component fermionic
coordinates.
The SUSY operators are $Q=\partial_\theta - \theta \partial_z$
and $\bar{Q}=\partial_{\bar{\theta}} - \bar{\theta} \partial_{\bar{z}}$.
The operators $D=\partial_\theta + \theta \partial_z$
and $\bar{D}=\partial_{\bar{\theta}} + \bar{\theta} \partial_{\bar{z}}$
commute with $Q$ and $\bar{Q}$.
Then,
the SUSY transformation of $ {\bf X}^\mu$ is explicitly given by
\begin{eqnarray}
 \delta X^\mu  &=& i \epsilon \psi^\mu + i \bar{\epsilon} \tilde{\psi}^\mu,
  \nonumber \\
 \delta \psi^\mu  &=& -i \epsilon \partial_z X^\mu 
- i \bar{\epsilon} F^\mu,
  \nonumber \\
 \delta \tilde{\psi}^\mu  &=& - i \bar{\epsilon} \partial_{\bar{z}} X^\mu
+i \epsilon F^\mu,
  \nonumber \\
 \delta F^\mu  &=& -i \epsilon \partial_z \tilde{\psi}^\mu
+i \bar{\epsilon} \partial_{\bar{z}} \psi^\mu,
\end{eqnarray}
where $z=\tau+i \sigma$ and $\tau$ is the periodic coordinate of $S^1$
of the cylinder
with the period $2 \pi$.
For the bulk 2d action, we can take general non-linear sigma model action
of $ {\bf X}^\mu$, but here for simplicity we consider a conventional superstring action
\begin{eqnarray}
S_{\rm worldsheet} = \frac{1}{4 \pi} \int dz d\bar{z}
\; \left[
\partial_{\bar{z}} X^\mu \partial_{z} X^\mu+F^\mu F^\mu
+\psi^\mu \partial_{\bar{z}} \psi^\mu 
+\tilde{\psi}^\mu \partial_{z} \tilde{\psi}^\mu
\right] \, .
\end{eqnarray}

At the boundary corresponding to the D-branes, 
a half of SUSY should be broken.
We will take the unbroken SUSY such that
$\epsilon=\bar{\epsilon}$.
For this SUSY, we have
\begin{eqnarray}
 \delta X^\mu  &=& i \epsilon (\psi^\mu + \tilde{\psi}^\mu),
  \nonumber \\
 \delta \psi^\mu  &=& -i \epsilon (\partial_z X^\mu + F^\mu),
  \nonumber \\
 \delta \tilde{\psi}^\mu  &=& - i \epsilon (\partial_{\bar{z}} X^\mu
-  F^\mu),
  \nonumber \\
 \delta F^\mu  &=& -i \epsilon (\partial_z \tilde{\psi}^\mu
- \partial_{\bar{z}} \psi^\mu).
\end{eqnarray}

A boundary condition which is consistent with this SUSY is
\begin{eqnarray}
 \partial_\sigma X^\mu|_{\rm b}=0, \,\,\, (\psi^\mu-\tilde{\psi}^\mu)|_{\rm b}=0, 
\,\,\,  \partial_\sigma (\psi^\mu+\tilde{\psi}^\mu)|_{\rm b}=0,  \,\,\,
F^\mu|_{\rm b}=0,
\label{susyb}
\end{eqnarray}
which is an off-shell extension of the Neumann boundary condition,
{\it i.e.} boundary condition for the D9-branes (and
anti-D9-branes).\footnote{
With the on-shell condition for the free bulk theory,
this set of the boundary conditions (\ref{susyb}) 
is equivalent to the usual Neumann boundary
condition of the superstring.
Instead of (\ref{susyb}), one can find a different set of boundary conditions
which is consistent with SUSY: $ \partial_\sigma X^\mu|_{\rm b}=0, \,\,\,  \partial_\sigma
(\psi^\mu+\tilde{\psi}^\mu)|_{\rm b}=0$.
If we further impose $ (\psi^\mu_0-\tilde{\psi}^\mu_0)|_{\rm b}=0$,
where $\psi_0 =\int d \tau \psi$, then this set of the  boundary conditions is
consistent with SUSY and is equivalent to the usual boundary condition
of the superstring at on-shell.
We are allowed to use this instead of (\ref{susyb}).}
At the boundary with this Neumann boundary condition,
the SUSY transformations for the non-zero fields are
\begin{eqnarray}
 \delta X^\mu  &=& i \epsilon \psi^\mu_{\rm b},
  \nonumber \\
 \delta \psi^\mu_{\rm b}  &=& -i \epsilon \partial_\tau X^\mu, 
\end{eqnarray}
where $\psi^\mu_{\rm b} = \psi^\mu+\tilde{\psi}^\mu$. 
Then, we can easily see that 
the boundary interactions constructed from the superfield including
\begin{eqnarray}
{\bf X}^\mu_{\rm b}\equiv X^\mu + i \theta_{\rm b} \psi^\mu_{\rm b} 
\label{superb}
\end{eqnarray}
and $D_b=\partial_{\theta_b} +\theta_b \partial_\tau$
are invariant under the SUSY
$Q_b=\partial_{\theta_b} -\theta_b \partial_\tau$.
Here the super-coordinates are $\tau$ and $\theta_{\rm b}=(\theta+\bar{\theta})/2$.
In this way, we can consider an arbitrary boundary interaction which preserves 
the half of the world sheet SUSY.

At the other boundary of the cylinder we will put
the following boundary condition which is 
consistent with the SUSY:
\begin{eqnarray}
0= \partial_z X^\mu_n|_{\rm b'}= \partial_{\bar z} X^\mu_{-n}|_{\rm b'}
=\partial_z (\psi^\mu_n+\tilde{\psi}^\mu_n)|_{\rm b'}
=\partial_{\bar z} (\psi^\mu_{-n}+\tilde{\psi}^\mu_{-n})|_{\rm b'},
\,\,\,\,\, {\rm for
 } \,\, n>0
\label{susyb2}
\end{eqnarray}
where $X^\mu_n \equiv \int e^{-in\tau} X^\mu$.
With the on-shell condition for the free theory, {\it i.e.} 
$0=\partial_z \partial_{\bar z} X
= \partial_z \tilde{\psi}=\partial_{\bar z} \psi=F$, 
we can see that
this boundary condition corresponds to the closed string vacua $|0 \ket_{\rm RR}$
in the RR sector
because the raising operators correspond to the positive modes of
$\partial_z (*)$
and the negative modes of
$\partial_{\bar z} (*)$.\footnote{
Instead this, we can also 
take 
$0= \partial_z X^\mu_n|_{\rm b'}= \partial_{\bar z} X^\mu_{-n}|_{\rm b'}
=\partial_z \psi^\mu_n|_{\rm b'}=\partial_z \tilde{\psi}^\mu_n|_{\rm b'}
=\partial_{\bar z} \psi^\mu_{-n}|_{\rm b'}
=\partial_{\bar z} \tilde{\psi}^\mu_{-n}|_{\rm b'}
=( ( \partial_z- \partial_{\bar z}) \partial_z X + 2 \partial_z F    )_n|_{\rm b'}
=( ( \partial_z- \partial_{\bar z}) \partial_{\bar z} X 
+ 2 \partial_z F    )_{-n}|_{\rm b'}
\,\,\,\,\, ({\rm for  } \,\, n>0)$ which is consistent with SUSY
and equivalent to the closed string vacua with the on-shell condition
for the free theory.
Note that the both of the boundary conditions are not consistent
with the SUSY with $\epsilon \neq \bar{\epsilon}$.
However, at the on-shell this corresponds to the closed string vacua,
thus there is no real problem and this will be an artifact 
of extending to the off-shell SUSY.
}
We can also insert 
the fermion zero modes $\psi_0^\mu =\frac{1}{4 \pi}
\int d \tau (\psi^\mu+\tilde{\psi}^\mu)$
which are SUSY invariants at the boundary.
Then, the insertion of the following
\begin{eqnarray}
-i \sum_{p={\rm odd}} 
\frac{1}{(p+1)!}
C^{(p+1)}_{\mu_0 \cdots \mu_p} 
(2i)^{-(p+1)/2}
\psi_0^{\mu_0}   \cdots  \psi_0^{\mu_p}
\end{eqnarray}
corresponds to the RR state $|C \ket_{\rm RR}$ where 
\begin{eqnarray}
C=\sum_{p={\rm odd}} C^{(p+1)} ,
\qquad
C^{(p+1)}\equiv \frac{ 1}{(p+1)!} 
C^{(p+1)}_{\mu_0 \cdots \mu_p} dx^{\mu_0} \cdots dx^{\mu_p}
\end{eqnarray}
is the background 
constant RR $(p+1)$-form
\cite{Kraus:2000nj}. 

Now our field theory on the cylinder 
is SUSY invariant, thus we can use the localization technique.
We define 
\begin{eqnarray}
 V &=& (\delta \psi^\mu)^\dagger \psi^\mu +
(\delta \tilde{\psi}^\mu)^\dagger \tilde{\psi}^\mu  \nonumber \\
&\equiv & i (\partial_{\bar{z}} X^\mu + F^\mu) \psi^\mu
+i (\partial_{z} X^\mu - F^\mu) \tilde{\psi}^\mu,
\end{eqnarray}
then $\delta \int d\tau d\sigma \delta V= \int d\tau d\sigma
\partial_\tau V=0$
and
\begin{eqnarray}
 \delta V= 2 \epsilon 
\left(
\partial_{\bar{z}} X^\mu \partial_{z} X^\mu+F^\mu F^\mu
+\psi^\mu \partial_{\bar{z}} \psi^\mu 
+\tilde{\psi}^\mu \partial_{z} \tilde{\psi}^\mu
+\frac{i}{2} \partial_\sigma (\psi^\mu \tilde{\psi}^\mu)
\right),
\end{eqnarray}
is the 
usual free massless action (where 
the last term vanish with the boundary conditions
and the saddle point equations).
By adding the regulator action
\begin{eqnarray}
S_{\rm reg}= t \int d\tau d\sigma\;  \delta V|_{\epsilon=1}
\label{regS}
\end{eqnarray}
and taking $t \rightarrow \infty$
limit,
we find saddle point equations
\begin{eqnarray}
 0=\partial_z X^\mu=\partial_{\bar{z}} X^\mu=F^\mu,
\end{eqnarray}
and 
\begin{eqnarray}
0=\partial_z \psi=\partial_{\bar{z}} \psi^\mu \, .
\label{fermionlocus}
\end{eqnarray}
The latter means that the non-zero modes of the fermions
can be put to be zero, since a rescaling of the fermions in the regulator
action (\ref{regS}) to have a canonical kinetic term has the 
equations of motion (\ref{fermionlocus}).

The 1-loop determinant around the locus is trivial because the zero modes do not 
couple to the non-zero modes in the regulator action.
Therefore, the supersymmetric localization tells us that in the evaluation of 
the partition function and physical observables we can drop the 
non-zero modes and can consider only the zero modes.

%%%%%%%%%%%%%%%%%%%%%%%%%%%%%%%%%%%%%%%%%%%%%%
\section{No creation of more D-branes}
\label{sec3}

%%%%%%%%%%%%%%%%%%%%%%%%%%%%%%%%%%%%%%%%
\subsection{The question}

Our interest here is whether one can climb up the potential hill in the BSFT, to obtain 
a solution of the BSFT representing multiple D-branes, starting from a single D-brane.
To be concrete, we consider only D9-branes as a starting point for the BSFT action. 
These include a BPS D9-brane,
and a pair of a D9-brane and an anti-D9-brane (brane-antibrane pair).

First, we need to summarize what we have seen for the localization. 
After the localization, the vacuum expectation value of any supersymmetric operator
at the worldsheet boundary 
can be evaluated at the localization locus, 
\begin{eqnarray}
\dot{X}^\mu(\tau)=0, \quad \dot{\psi}^\mu(\tau)=0 \, .
\label{susylocus}
\end{eqnarray}
The dot denotes a derivative with respective to the boundary coordinate $\tau$
of the worldsheet.
$\mu$ runs from 0 to 9
labeling the target spacetime dimensions. 
In the following we omit the suffix  ``${\rm b}$" 
for ${\bf X}_{\rm b}$, $\theta_{\rm b}$ and
$\psi_{\rm b}$ in (\ref{superb}).

The generic boundary interaction is written in terms of the superfield in 1 dimension
(the boundary of the worldsheet),
\begin{eqnarray}
{\bf X}^\mu \equiv X^\mu(\tau)  + 
i \theta \psi^\mu(\tau) \, ,
\end{eqnarray}
and we allow arbitrary supersymmetric vertex operators, as BSFT is defined 
as a complete set of them.

Our aim is to show whether multiple D-branes can be created in the BSFT or not.
We already know that the tachyon condensation can make a D-brane vanish,
but the problem here is to see whether one can create one more unit of the  D-brane
charge by a condensation of some massive modes of string theory.
See Fig.~\ref{fig}.
We define ``seeing the creation" by looking at the total D-brane charge. 
So, for example, creation of a brane-antibrane pair cannot be detected by
our formalism.

Of course, it is easy to create a charge of a D-brane with different spatial world volume 
dimensions. 
For example, it is well-known that, 
on a BPS  D-brane, turning on a constant magnetic field results in a different 
D-brane charge (of a D-brane with lower dimensions). 
So the question of our concern is: let us start with a D9-brane,
and with a condensation of massive modes on the D9-brane, can we create
a charge of multiple D9-branes?

%%%%%%%%%%%%%%%%%%%%%%%
\FIGURE[t] {
\includegraphics[width=13cm,bb=0 0 720 540]{for-paper.pdf}
\caption{
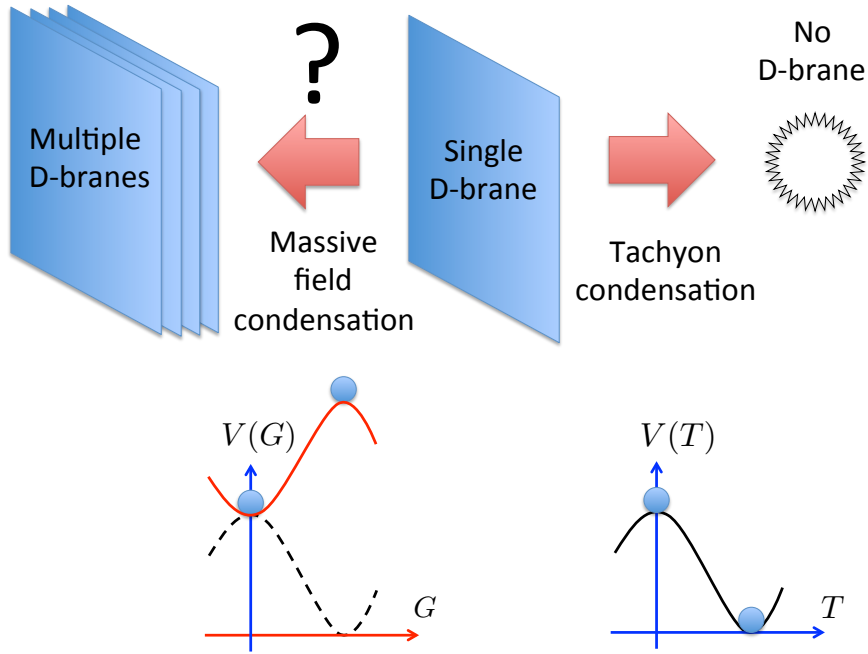
A schematic picture of the question of the D-brane creation.
For the right part, a popular (and proven) D-brane annihilation by
the tachyon condensation is shown. The left part is our question.
To create multiple D-branes, more energy is necessary for the 
D-brane tensions, thus a condensation of massive modes $G$ 
which climbs up
the potential hill would be naturally expected. 
}
\label{fig}
}
%%%%%%%%%%%%%%%%%%%%%%%

There are two possible starting points. The first one is a single BPS D9-brane.
The second one is a pair of BPS D9-brane and anti-D9-brane. The latter
was used for the tachyon condensation giving the annihilation of the pair 
\cite{Takayanagi:2000rz}.  But in fact, for our purpose,
nothing can prevent us from analyzing the first case. We check whether 
string theory can have a power to obtain multiple D-brane charges
starting from the (naive) Hilbert space of a single D-brane.

%%%%%%%%%
\subsection{BSFT for a BPS D9-brane}

First, let us investigate the case of a single BPS D9-brane. The question is ---
can we create another D9-brane by making a condensation of massive modes on
the BPS D9-brane?

It would be instructive to review what would happen to the 
condensation of a massless mode. Later we will consider a massive mode,
and study the generic case.
The massless mode of an open string is the photon vertex operator,
\begin{eqnarray}
I_{\rm B} = \int \! d\tau d\theta \;
\left(
-i D_\theta{\bf X}^\mu
A_{\mu}[{\bf X}]
\right) \, .
\label{photonv}
\end{eqnarray}
If we make an integration of the boundary $\theta$ coordinate, we obtain
\begin{eqnarray}
I_{\rm B} = \int \! d\tau  \;
\left(
-i \dot{X}^\mu
A_{\mu}[X]
+  \frac{i}{2} \psi^\mu \psi^\nu
F_{\mu\nu}[X] \right) \, .
\end{eqnarray}
The BSFT action for the RR sector 
is nothing but the partition function of the worldsheet
theory
with some fermion zero modes corresponding to the background RR-forms,
so basically we evaluate the expectation value of the operator $\exp[-I_{\rm B}]$.
When evaluating the expectation value (VEV), we use the localization. The locus satisfies
(\ref{susylocus}), so we substitute (\ref{susylocus}) into the boundary interaction, to have
\begin{eqnarray}
\bra e^{-I_{\rm B}}\ket  = 
{\cal N} \exp \left[
-\frac{i}{2}\int \! d\tau  \;
\psi_0^\mu \psi_0^\nu
F_{\mu\nu}[X_0]  
\right] \, .
\label{Fpp}
\end{eqnarray}
Note that the VEV $\bra\; \ket$ is defined for massive modes, and the zero modes
$\psi_0$ and $X_0$ are not integrated yet. The overall constant ${\cal N}$ is can be evaluated
by the localization explicitly but here we do not need it, as it will be a normalization of
the RR charge.

The expression (\ref{Fpp}) is a well-known formula for the RR charge.
If the field strength $F_{\mu\nu}$ of the photon is constant, the expansion of 
the exponential form of (\ref{Fpp})
in terms of the field strength supplies multiple fermionic zero modes. Those zero modes
are cancelled precisely by the RR vertex insertion, 
so we obtain lower dimensional D-brane
charges. Note here that after the expansion of the exponential, the first term 
(the term without the fermion zero mode) is of course the unity. It means that the boundary
state given by this boundary deformation has a unit RR charge for the D9-brane,
in addition to the lower dimensional D-brane charges. 
So, we see here that 
the photon condensation can never give a RR charge of multiple D9-branes 
if we start from 
a D9-brane.
The condensation is accompanied with fermionic zero modes $\psi_0$ in (\ref{Fpp}),
thus creating only  lower dimensional D-brane charges.

Next, let us visit an example of a  concrete massive state. We consider 
a vertex for the first massive state on the BPS D9-brane
(see for example \cite{Hashimoto:2004qp}). The vertex
is represented by a boundary action of the worldsheet,
\begin{eqnarray}
I_{\rm B} = \int \! d\tau d\theta \;
\left(
D_\theta{\bf X}^\mu
D_\theta{\bf X}^\nu
D_\theta{\bf X}^\rho
V_{\mu\nu\rho}[{\bf X}]
\right.
\hspace*{30mm}
\nonumber \\
\left.
+D^2_\theta{\bf X}^\mu
D_\theta{\bf X}^\nu
W_{\mu\nu}[{\bf X}]
+D^3_\theta{\bf X}^\mu
S_{\mu}[{\bf X}]
\right) \, .
\label{VWS}
\end{eqnarray}
Here normally the arbitrary function $V, W$ and $S$ can be expanded
by plane waves, and the indices $\mu, \nu$ and $\rho$ provides
the polarization of the states. One can show that $S_\mu$ and
the anti-symmetric part of $W_{\mu\nu}$ can be gauged away. 
So we need to consider only the symmetric part of $W_{\mu\nu}$
and the antisymmetric $V_{\mu\nu\rho}$.

Since the 
expression $I_{\rm B}$ is explicitly written only by the super field ${\bf X}$,
the state is supersymmetric. After making the integration of the boundary
fermion coordinate $\theta$, one obtains 
$I_{\rm B} \equiv I_{\rm B}[W] + I_{\rm B}[V]$ with 
\cite{Hashimoto:2004qp}
\begin{eqnarray}
I_{\rm B}[W] &=& \int \! d\tau \;
\left[
 \dot{X}^\mu \psi^\nu \psi^\rho \partial_\rho W_{\mu\nu}(X) 
+(\dot{X}^\mu \dot{X}^\nu - \dot{\psi}^\mu \psi^\nu) W_{\mu\nu}(X)
\right] \, , \quad \\
I_{\rm B}[V] &=& \int \! d\tau \;
\left[
 \psi^\mu \psi^\nu \psi^\rho \psi^\sigma \partial_\sigma V_{\mu\nu\rho}(X)
-3  \psi^\mu \psi^\nu \dot{X}^\rho V_{\mu\nu\rho}(X)
\right] \, .
\end{eqnarray}

Now, let us consider a condensation of the massive mode. This means
that we have 
\begin{eqnarray}
W_{(\mu\nu)}(X_0) \neq 0 \, , \quad
V_{[\mu\nu\rho]} (X_0)\neq 0 \, .
\end{eqnarray}
Our interest is, for this condensation, whether we can 
have a RR-charge of the D9-brane or not. To evaluate the 
VEV of $e^{-I_{\rm B}}$, let us substitute the localization locus
condition (\ref{susylocus}) into the boundary action. We obtain
\begin{eqnarray}
I_{\rm B}\bigg|_{\rm locus} &=& \int \! d\tau \;
 \psi_0^\mu \psi_0^\nu \psi_0^\rho \psi_0^\sigma \partial_\sigma V_{\mu\nu\rho}(X_0)
\, . \quad 
\end{eqnarray}
It is obvious that this result has the same property as the case of the photon (\ref{Fpp}).
All the condensation fields are accompanied by the fermion zero modes $\psi_0$,
so they merely gives lower-dimensional D-brane charges. They never create the
additional D9-brane charge.

After examining some examples, we can come to a generic statement. On the BPS D9-brane, generic boundary action can be written as
\begin{eqnarray}
I_{\rm B} = \sum_{n_1,n_2,\cdots}\int \! d\tau d\theta \;
f_{n_1,n_2,\cdots}[{\bf X}]
(D_\theta{\bf X})^{n_1}
((D_\theta)^2{\bf X})^{n_2}
\cdots 
((D_\theta)^k {\bf X})^{n_k}
\cdots 
\, .
\end{eqnarray}
Using the component expression, we find
\begin{eqnarray}
D_\theta {\bf X} =i \psi + \theta \dot{X}, \quad
D^2_\theta {\bf X} = \dot{X} + i \theta \dot{\psi}, \quad \cdots 
\end{eqnarray}
and 
\begin{eqnarray}
f[{\bf X}] = f(X) + i \theta \psi \partial f[X],
\end{eqnarray}
Then the boundary integral $d\theta$
is performed, under the condition of the localization locus (\ref{susylocus}), to give
\begin{eqnarray}
I_{\rm B}\bigg|_{\rm locus} = 
\sum_{n_1}\int \! d\tau \;
\partial g_{n_1}[X_0]
(\psi_0)^{n_1+1}
\, ,
\label{IBr}
\end{eqnarray}
where $g_{n_1}\equiv f_{n_1,n_2=n_3=n_4=\cdots=0}$. This is because
only $D_\theta {\bf X}$ has a component without the time derivative,
and the other $(D_\theta)^k {\bf X}$ with $k\geq 2$ vanishes due to (\ref{susylocus}).
Since $n_1$ is a non-negative integer, the resultant (\ref{IBr}) includes at least one
fermionic zero mode $\psi_0$. Therefore, any condensation gives only lower-dimensional RR charges. This concludes a proof that, on a BPS D9-brane, condensation of any
open string massive mode cannot change the D9-brane RR charge. Hence the 
BSFT cannot
accommodate a D-brane creation starting from a single BPS D-brane.

\subsection{BSFT for a D9-antiD9 pair}  

For the case of a pair of a D9-brane and an anti-D9-brane, we can generalize the
proof found in the previous subsection. The BSFT action for a brane-antibrane was 
found in \cite{Takayanagi:2000rz} and \cite{Kraus:2000nj}. 
For the boundary interaction, the difference from 
the BPS D-brane studied above is just the inclusion of a fermionic boundary superfield
\begin{eqnarray}
{\bf \Gamma} = \eta(\tau) + \theta F(\tau) \, .
\end{eqnarray}
For the brane-antibrane pair, this field is complex, while for a non-BPS D-brane,
this field is taken to be real \cite{Kutasov:2000aq} (see also \cite{Harvey:2000na,Tseytlin:2000mt}). 

Originally the boundary field $\eta$ was introduced by Witten \cite{Witten:1998cd} for the purpose of giving a Chan-Paton factor to represent the non-BPS D-brane before a worldsheet projection. So the inclusion of this ${\bf \Gamma}$ in the boundary interaction
is basically just linear. The field $F(\tau)$ is an auxiliary field, since the kinetic term
for this boundary superfield is
\begin{eqnarray}
S_{\bf \Gamma} \equiv - \int \! d\tau d\theta \; \bar{\bf \Gamma} D_\theta {\bf \Gamma}
= \int \! d\tau \left[\bar{\eta} \dot{\eta} - \bar{F}F \right]
\, .
\end{eqnarray}

Now, let us write a generic boundary interaction as before, with a possible linear 
component in ${\bf \Gamma}$, as\footnote{On the brane-antibrane, there are vertex operators without the boundary super field ${\bf \Gamma}$. For that vertices,
the discussion reduces to that of the previous subsection, so we do not treat those
in the following.}
\begin{eqnarray}
I_{\rm B}=\sum_{n_1,n_2,\cdots}\int \! d\tau d\theta \;
{\bf \Gamma}\,
g_{n_1,n_2,\cdots}[{\bf X}]
(D_\theta{\bf X})^{n_1}
(D^2_\theta{\bf X})^{n_2}
\cdots 
(D^k_\theta{\bf X})^{n_k}
\cdots 
+ \,\,\, {\rm c.c.}
\nonumber \\
+ 
\sum_{n_1,n_2,\cdots}\int \! d\tau d\theta \;
{\bf \Gamma} \bar {\bf \Gamma}\,
G_{n_1,n_2,\cdots}[{\bf X}]
(D_\theta{\bf X})^{n_1}
(D^2_\theta{\bf X})^{n_2}
\cdots 
(D^k_\theta{\bf X})^{n_k}
\cdots 
\, . \quad 
\label{generalT}
\end{eqnarray}
The new part is ${\bf \Gamma}$, $\bar{\bf \Gamma}$ and ${\bf \Gamma}\bar{\bf \Gamma}$. 

A popular example is a tachyon vertex operator. This corresponds to having
$n_1=n_2=\cdots = 0$ for the first line in (\ref{generalT}),
\begin{eqnarray}
I_{\rm B}[T] = \int \! d\tau d\theta \; \frac{1}{\sqrt{2\pi}} 
\left(
T[{\bf X}] \bar{\bf \Gamma} +\bar{T}[{\bf X}] {\bf \Gamma}
\right)
 \, .
 \label{tachyon}
\end{eqnarray}
So the general expression (\ref{generalT}) includes the tachyon condensation
as a particular case. The massless gauge fields on each D9-brane are also included.

Now, the localization locus condition (\ref{susylocus}) is applied to the vertex
operator (\ref{generalT}) as before, to obtain the following expression
\begin{eqnarray}
I_{\rm B}\bigg|_{\rm locus} &  = & 
\sum_{n_1}\int \! d\tau 
\left(
\partial g_{n_1}[X_0]
(i \psi_0)^{n_1+1} \eta
+ 
g_{n_1}[X_0]
(i \psi_0)^{n_1} F
\right)
+ \,\,\, {\rm c.c.}
\nonumber \\
& & + 
\sum_{n_1}\int \! d\tau 
\left(
\partial G_{n_1}[X_0]
(i \psi_0)^{n_1+1} \eta \bar{\eta}
+ 
G_{n_1}[X_0]
(i \psi_0)^{n_1} (F\bar{\eta} + \bar{F}\eta)
\right) \, .
\nonumber \\
\end{eqnarray}
Since we are interested in a D9-brane charge, only the choice $n_1=0$ is a candidate. 
(Other value of $n_1$ provides a creation of lower dimensional D-brane charge.)
So dropping other terms gives\footnote{In the RR sector of the BSFT action,
we need a supertrace, i.e. a factor $[\bar{\eta},\eta]$ in the path integral measure 
for the $\eta$ and $\bar{\eta}$ integration \cite{Takayanagi:2000rz}.}
\begin{eqnarray}
I_{\rm B}\bigg|_{\rm locus} = 
\sum_{n_1}\int \! d\tau 
\left( 
g_{n_1=0}[X_0] F
 + \,\, {\rm c.c.}
+
G_{n_1=0}[X_0] (F\bar{\eta} + \bar{F}\eta)
\right) \, .
\quad 
\label{IBr3}
\end{eqnarray}
This term would be a possible term which can create the D9-brane charge.
However, the first term is nothing but the tachyon coupling (\ref{tachyon}). 
The second term can be generated by a field redefinition of the tachyon $T[X]$ 
(see the discussion in section 2 in \cite{Takayanagi:2000rz}). 
So, this (\ref{IBr3}) is not a massive excitation of the open string --- it is merely 
a tachyon coupling, which has been already shown to be unable to produce
an additional  D9-brane RR charge.\footnote{Since we assume the homogeneity for the D9-brane,
the field $g_{n_1=0}[X_0]$ needs to be a constant.}

This concludes a proof that a D9-brane charge cannot be generated by any condensation
of massive excitation modes of open superstring in BSFT.

%%%%%%%%%%%%%%%%%%%%%%%%%%%%%%%%%%%%%%%%

\section{General RR-coupling formula}
\label{sec4}

In the previous section, we have shown that the D-brane change cannot be created
by any condensation of massive open strings. This applies only to the charge of the same
kind of D-branes: In our previous cases those are D9-brane charges.
On the other hand, lower-dimensional D-brane charges can be easily created.
Here we shall derive a generic RR-charge formula including any massive mode condensation,
which generalizes the known RR charge formula originally written 
only with condensation of 
the massless and the tachyonic modes on the D-brane(s).

\subsection{General RR-coupling formula for a BPS D9-brane}

For the massless mode condensation the popular expression for the RR coupling
in the D-brane action is the Chern-Simons term (or often called ``Wess-Zumino term"),
\begin{eqnarray}
S_{\rm RR}= T_{D9} \sum_{p \; : \;  {\rm odd}}  \int  C^{(p+1)}
 \wedge e^{2 \pi \alpha' F} 
\label{CSterm}
\end{eqnarray}
where $C^{(p+1)}$ with an odd integer $p$ is the RR $(p+1)$-form, 
and $F$ is the world volume
gauge field strength which is 2-form. The integrand in (\ref{CSterm}) is chosen 
in such a way that the total degree of the form is equal to the world volume dimension of
the D-brane (which is 10 for the case of the BPS D9-brane).
In the course of generalizing this formula, we just look back how this (\ref{CSterm}) 
was derived. The boundary action (\ref{Fpp}) is precisely the origin of the formula.
A nonzero scattering amplitude after the path integral of the fermion zero mode $\psi_0$
requires an insertion of the RR vertex
\begin{eqnarray}
C^{(p+1)}_{\mu_0\mu_2\cdots\mu_p} \psi_0^{\mu_0}\cdots\psi_0^{\mu_p}
\end{eqnarray}
so that the total number of the fermion zero mode is 10 (and completely antisymmetric
under the exchange in variables $\mu_i$).

Now, let us generalize the formula to include a condensation of the open string massive modes. The result of the localization for the worldsheet boundary action is already given by (\ref{IBr}).
More specifically, the nonzero contribution to $I_{\rm B}$ after the localization comes only
from a specific type of the open string excitation
\begin{eqnarray}
I_{\rm B} = -i\int\! d\tau d\theta \; g_{\mu_1\cdots \mu_n}^{(n)}[{\bf X}]D_\theta {\bf X}^\mu_1 \cdots D_\theta {\bf X}^{\mu_n}\, ,
\end{eqnarray}
where the indices $\mu_1,\cdots,\mu_n$ are mutually anti-symmetric. The integer $n$ should be an odd integer for a BPS D9-brane, due to the GSO projection. The $n=1$ case corresponds to
the massless gauge field (\ref{photonv}). The $n=3$ case is a part of the first excited massive mode (\ref{VWS}). Following the same logic for the fermion zero mode integration, we arrive at a generic formula
\begin{eqnarray}
S_{\rm RR}= T_{D9}
\sum_{p \; : \;  {\rm odd}}  \int  C^{(p+1)} \wedge \exp \left[4\pi \sum_{m=1}^{5} dg^{(2m-1)}
\right] \, .  
\label{CSterm2}
\end{eqnarray}
Here in the exponent, $dg^{(2m-1)}$ is the $2m$-form field strength of the
open string massive mode $g^{(2m-1)}(x)$,  
\begin{eqnarray}
g^{(2m-1)} 
=(-2i)^{m-1}
g_{\mu_1\cdots \mu_{2m-1}}^{(2m-1)}(x) \,
dx^{\mu_1}\wedge \cdots \wedge dx^{\mu_{2m-1}} \, .
\end{eqnarray}
%In the formula, 
%we have neglected the numerical factors.
%They will be fixed in the next subsection.
The numerical factors in the formula 
will be determined in a more general study in the next subsection.

The important point is that the RR coupling formula 
(\ref{CSterm2}) is for {\it all} open string excitations. 
The open string fields appearing in the formula, $g^{(2m-1)}$ with $m=1,2,3,4,5$ 
are the only fields which can contribute to the RR charge. In particular,
we have found that the open string excitations which can contribute to
the RR charge is restricted to mass level 4.

\subsection{General RR-coupling formula for multiple D9-branes 
}

Next, we apply the same strategy for the D9-antiD9 pairs. We shall find that
the most general RR coupling formula is written by the 
Quillen's superconnection 
including higher form fields.

Let us consider $N$ D9-D9bar branes with $N=2^{n-1}$.
This system is realized by introducing the 
boundary auxiliary superfields 
${\bf \Gamma}^i$ 
where $i=1,\ldots, 2n$.\footnote{
A generalization to the non BPS D9-branes in type IIA superstring theory is 
easily achieved by taking $i=1,\ldots, 2n+1$. 
Here we have moved to another notation for the gamma matrices for
convenience. For $n=2$,
they are related to the previous ones as 
${\bf \Gamma} = (1/2)({\bf \Gamma}^1 + i {\bf \Gamma}^2)$ 
and $\bar{\bf \Gamma} = (1/2)({\bf \Gamma}^1 - i {\bf \Gamma}^2)$.
}
The action with general fluctuations at the boundary is 
\begin{eqnarray}
 -\int d \tau d \theta \frac{1}{4}
({\bf \Gamma}^i D_\theta {\bf \Gamma}^i)
+I_{\rm B} ({\bf \Gamma}, 
{\bf X}),
\end{eqnarray}
where $I_{\rm B}$ is a general boundary action. 
The boundary action $I_{\rm B}$ does not include 
$D_\theta {\bf \Gamma}$ 
because
${\bf \Gamma}$ 
is an auxiliary field.
Thus we can write $I_{\rm B}$ as
\begin{eqnarray}
 I_{\rm B}= -i \int d \tau d \theta \, A_{n_1,n_2,\ldots ; m}({\bf X})
(D_\theta {\bf X})^{n_1}  ((D_\theta)^2 {\bf X})^{n_2} \cdots \times
({\bf \Gamma})^m, 
\end{eqnarray}
where 
\begin{eqnarray}
&&
  A_{n_1,n_2,\ldots ; m}({\bf X})
(D_\theta {\bf X})^{n_1}  ((D_\theta)^2 {\bf X})^{n_2} \cdots \times ({\bf \Gamma})^m
\nonumber
\\
&&
\qquad
=A_{ (\mu_1^1,\mu_2^1,\cdots, \mu_{n_1}^1 )  ,
(\mu_1^2,\mu_2^2,\cdots, \mu_{n_2}^2 ), \cdots;
(k_1,k_2,\cdots,k_m)}({\bf X})
\nonumber \\
&& 
\qquad\qquad
\times (D_\theta {\bf X}^{\mu_1^1}  D_\theta {\bf X}^{\mu_2^1}  \cdots D_\theta
X^{\mu_{n_1}^1} ) 
\nonumber \\
&& 
\qquad\qquad
\times
((D_\theta)^2 X^{\mu_1^2}  (D_\theta)^2 X^{\mu_2^2}  \cdots (D_\theta)^2
X^{\mu_{n_2}^2} ) \cdots 
\nonumber \\
&& 
\qquad\qquad
\times
{\bf \Gamma}^{k_1} {\bf \Gamma}^{k_2} \cdots {\bf \Gamma}^{k_m}.
\end{eqnarray}
Here, $I_{\rm B}$ needs to be bosonic, thus $A=0$ if $\sum_{a={\rm odd}}n_a+m=2
{\mathbf Z}+1$. This condition is nothing but the GSO projection.
We also impose the reality condition: ${I_{\rm B}}^*=-I_{\rm B}$
in the Wick rotated world sheet action.

We shall check infinitesimal gauge transformations to find a consistent 
non-Abelian nature of the boundary interaction written by ${\bf \Gamma}^i$.
The gauge transformations are 
generated by adding the following general total divergence term to the boundary action,
\begin{eqnarray}
 0 &=-i & \int d \tau d \theta D_\theta \left( 
\lambda_{n_1,n_2,\cdots;m} ({\bf X})
(D_\theta {\bf X})^{n_1}  (D_\theta {\bf X})^{n_2} \cdots \times ({\bf \Gamma})^m
\right)
\nonumber \\
&=&
-i \int d \tau d \theta \Bigg(
\left(
\frac{\partial}{\partial X^\mu} \lambda_{n_1,n_2,\cdots;m}
\right)
D_\theta {\bf X}^\mu (D_\theta {\bf X})^{n_1} \cdots ({\bf \Gamma})^m
\nonumber 
\\
&&
\qquad\qquad\qquad
+  \lambda_{n_1,n_2,\cdots;m} D_\theta 
\left(
(D_\theta {\bf X})^{n_1} \cdots \right) ({\bf \Gamma})^m
\nonumber 
\\
&& 
\qquad\qquad\qquad
+(-1)^m  \lambda_{n_1,n_2,\cdots;m} 
(D_\theta {\bf X})^{n_1}  (D_\theta {\bf X})^{n_2} \cdots \times D_\theta ({\bf \Gamma})^m
\Bigg),
\end{eqnarray}
where $\lambda_{n_1,n_2,\cdots;m}$ are the gauge transformation
parameters with the appropriate reality conditions
and the following condition:
$\lambda=0$ if $\sum_{a={\rm odd}}n_a+m=2{\mathbf Z}$. 
The last term has $D_\theta$ acting on ${\bf \Gamma}$. 
In order to maintain 
the usual kinetic term for the auxiliary superfields ${\bf \Gamma}$,
we need to redefine them as follows:
\begin{eqnarray}
{\bf \Gamma'}^i = {\bf \Gamma}^i 
+2 i \frac{\partial}{\partial 
 {\bf \Gamma}^i  }
\left(
\lambda_{n_1,n_2,\cdots;m} ({\bf X})
(D_\theta {\bf X})^{n_1}  (D_\theta {\bf X})^{n_2} \cdots \times ({\bf \Gamma})^m 
\right).
\end{eqnarray}
This field redefinition induces the following terms which are linear in $\lambda$:
\begin{eqnarray}
-2 i  
\left( \frac{\partial}{\partial  {\bf \Gamma}_i  } A \right)
\left( \frac{\partial}{\partial  {\bf \Gamma}_i  }\lambda \right).
\label{naive}
\end{eqnarray}
However, this is a naive expression because
we need to consider composite operators by 
the nonzero correlators between $\Gamma$'s.
Taking this effect into account, instead of (\ref{naive}),
we will have 
\begin{eqnarray}
 i [\lambda,A]=i (\lambda *A -A * \lambda),
\end{eqnarray}
where $*$ represents the fermionic $*$-product
defined in \cite{AST}.
Thus, the gauge transformation is given by
\begin{eqnarray}
  A \rightarrow A+d \lambda +i [\lambda,A]+e(\lambda),
\end{eqnarray}
where 
\begin{eqnarray}
 e(\lambda) = \lambda_{n_1,n_2,\cdots;m} D_\theta 
\left(
(D_\theta {\bf X})^{n_1} \cdots \right) ({\bf \Gamma})^m.
\end{eqnarray}

By the localization, only the fields with $n_a=0$ ($a>1$)
remain. 
The remaining fields precisely form the Quillen's 
superconnection ${\cal A}$.
We will represent it as follows:
\begin{eqnarray}
 \mca=\mca_{\mu_1,\cdots, \mu_n}^{k_1,\cdots, k_m} 
dx^{\mu_1} \cdots dx^{\mu_n} \gamma^{k_1} \cdots \gamma^{k_m},
\end{eqnarray}
where we have replaced ${\bf \Gamma}^k$ to $\gamma^k$ which is the gamma
matrix and $D_\theta {\bf X}^\mu$ to $dx^\mu$.
The gauge transformation parameter is also given by
\begin{eqnarray}
\lambda=\lambda_{\mu_1,\cdots, \mu_n}^{k_1,\cdots, k_m} 
dx^{\mu_1} \cdots dx^{\mu_n} \gamma^{k_1} \cdots \gamma^{k_m}. 
\end{eqnarray}
Then, the gauge transformation for the remaining fields is given by
\begin{eqnarray}
  {\cal A} \rightarrow {\cal A}+d \lambda +i [\lambda, {\cal A}].
\end{eqnarray}
Here the commutator using the fermionic $*$-product
is identified as the supercommutator.
The field strength defined by 
\begin{eqnarray}
{\cal F}=i (D_{\cal A})^2=d{\cal A}-i {\cal A}^2,
\end{eqnarray}
where $D_{\cal A}\equiv d -i {\cal A}$, and ${\cal F}$ is transformed as
${\cal F} \rightarrow {\cal F}+i [\lambda,{\cal F}]$.
We see that ${\rm Str} (f({\cal F}))$ is gauge invariant where ${\rm Str}$ is the
supertrace for the superconnection.

Now, from the result for the BPS D-brane,
it is almost clear that 
the cylinder partition function, which is 
the RR coupling of the D-branes
and should be gauge-invariant by definition, is given by
\begin{eqnarray}
 Z = 2^{-5}\,{\cal N} \int \sum_{p=odd} (-2 i)^{\frac{9-p}{2} } C^{(p+1)} \, \wedge {\rm Str} (e^{2 \pi i \cal F}),
 \label{generalRR}
\end{eqnarray}
where we have replaced $\psi_0^\mu$ by $(-i)dx^\mu$, and 
${\cal N}$ is the overall normalization, which will be fixed to ${\cal N}=2^5 \,T_{D9}$.  
This (\ref{generalRR}) is written by the super field strength of the superconnection,
and our final result for the general Ramond-Ramond coupling of the D9-antiD9 branes.

We can derive the RR coupling (\ref{generalRR}) explicitly, as follows.
The boundary action which will survive after the localization can be
written as
\begin{eqnarray}
I_{\rm B}=-\int\!\rmd\tau\rmd\theta \Bigl[\frac{1}{4}\bfg^I D_\theta \bfg^I 
+i \sum_{m=0}^{2n} 
{\cal A}^{I_1\cdots I_m}({\bf X}, D_\theta {\bf X}) \, 
\bfg^{I_1} \cdots \bfg^{I_m}\Bigr].
\end{eqnarray}
For the evaluation of the ${\bf \Gamma}$ integral, 
we can use the identity shown in \cite{AST}
\begin{eqnarray}
 \int D {\bf \Gamma}  \; e^{
\int\!\rmd\tau\rmd\theta \Bigl[\frac{1}{4}\bfg^I D_\theta \bfg^I 
+ {\bf M(\Gamma) } \Bigr]
}=\mathrm{Str} \, {\rm P} \,e^{\int d
  \tau (M_1(\gamma)-(M_0(\gamma))^2)},
\label{aa}
\end{eqnarray}
where 
${\bf M({\bf \Gamma})}=M_0({\bf \Gamma}) + \theta M_1({\bf \Gamma}) $
in which only the superfield ${\bf X}$ was decomposed to the component fields.
Here the r.h.s of (\ref{aa}) is represented by the 
corresponding gamma matrix
and ${\rm P}$ represents the path-ordering.
Using this with the zero mode reduction by the localization,
we have
$M(\gamma)=i \mca (X_0+i \theta \psi_0,i \psi_0)=i \mca (X_0,i
\psi_0)-\theta \psi_0^\mu \frac{\partial }{\partial X_0^\mu} \mca (X_0,i \psi_0)$
and
\begin{eqnarray}
\int\!D {\bf \Gamma} \; e^{-I_{\rm B}}=\mathrm{Str}\,e^{2\pi i\mcf}. 
\end{eqnarray}
Therefore the general RR coupling formula is given by (\ref{generalRR}).

The expression (\ref{generalRR}) is consistent with the charge quantization, 
because the integral of Chern character of the super connection, ${\rm Str} \exp({{\frac{ \cal F}{2 \pi}}})$,
will be quantized. With the help of the relation among D-brane tensions 
$T_{D(p-2)}/T_{Dp}=8 \pi^2$, one can show that the D-brane charge is quantized 
with our general result (\ref{generalRR}).

In the RR coupling formula (\ref{generalRR}) the terms appearing in the exponent
may look different from the standard normalization. This is just due to the convention.
If we replace $\psi_0^\mu$ by $ \sqrt{2 i} \, dx^\mu$ (instead of the previous $(-i)dx^\mu$),
the RR coupling is rewritten by a formula with a more familiar Chern-Simons couplings as
\begin{eqnarray}
 S_{\rm RR} (=Z) = T_{D9}  \int C \, \wedge {\rm Str} (e^{4 \pi  \tilde{\cal F}}),
 \label{generalRR2}
\end{eqnarray}
where 
\begin{eqnarray}
 \tilde{\cal F} \equiv \sum_n (-2i)^{ \frac{n-1}{2} } 
{\cal F}^{(n+1)}.
\end{eqnarray}
Here $ {\cal F} =\sum_n {\cal F}^{(n+1)}$ and ${\cal F}^{(n+1)}$
is an $(n+1)$-form.
Note that this replacement of $dx^\mu$
does not affect the result except for a change of the overall constant,
because (\ref{generalRR}) contains the volume form only.
In this formula we can see that the overall normalization is correct.

This $\tilde{\mcf}$ can also be considered as the field strength of a 
supperconnection.
Indeed, with
\begin{eqnarray}
 \tilde{\mca}=\sum_n (-2 i)^{\frac{n-1}{2}} \mca_n, \,\,\,\,\,
 \tilde{\lambda}=\sum_n (-2 i)^{\frac{n}{2}} \lambda_n,
\end{eqnarray}
where $\mca=\sum_n \mca_n, \,\,\, \lambda=\sum_n \lambda_n$,
we have $\tilde{\mcf}=i (d-i\tilde{\mca})^2$
and the gauge transformation is given by
$\tilde{\mca} \rightarrow \tilde{\mca} +d \tilde{\lambda} 
+ i [\tilde{\lambda},\tilde{\mca}]$.

%%%%%%%%%%%%%%

Let us examine the reality condition.
First, from the reality condition of the boundary action,  
we find a Hermiticity condition
\begin{eqnarray}
 (\mca_{\mu_1,\cdots, \mu_n}^{k_1,\cdots, k_m})^\dagger =
(-1)^{n+1+\frac{n+m-1}{2} } (\mca_{\mu_1,\cdots, \mu_n}^{k_1,\cdots, k_m}),
\label{Hermite}
\end{eqnarray}
where we have used $n+m=2 {\bf Z}+1$ and also the following manipulation
\begin{eqnarray}
 (i \mca)^\dagger=(-1)^{n+1+\frac{(n+m-1)(n+m)}{2} }  i (\mca_{\mu_1,\cdots, \mu_n}^{k_1,\cdots, k_m})^\dagger 
D_\theta {\bf X}^{\mu_1} \cdots D_\theta{\bf X}^{\mu_n} {\bf \Gamma}^{k_1}
\cdots {\bf \Gamma}^{k_m}.
\end{eqnarray}
Now we consider the p-form valued $2^n \times 2^n$ matrix
\begin{eqnarray}
 \mca=\mca_{\mu_1,\cdots, \mu_n}^{k_1,\cdots, k_m} ({X})
dx^{\mu_1} \cdots dx^{\mu_n} \gamma^{k_1} \cdots \gamma^{k_m}
= \begin{pmatrix}
A^+({X}) & -i \bar{T}({X})\\
-i T({X}) & A^-({X})  
\end{pmatrix}, 
\label{gammaAT}
\end{eqnarray}
where in the r.h.s. of the equation we decomposed it
to $2^{n-1} \times 2^{n-1}$ matrices.
With this decomposition,
we find\footnote{
Let us explain how we take the conjugate of (\ref{gammaAT}). We have
$(\gamma^{k_1} \cdots \gamma^{k_m})^\dagger 
= \gamma^{k_1} \cdots \gamma^{k_m} (-1)^{m(m-1)/2}$ and 
$ (dx^{\mu_1} \cdots dx^{\mu_n})^\dagger = (dx^{\mu_n} \cdots
dx^{\mu_1}) =(dx^{\mu_1} \cdots dx^{\mu_n}) (-1)^{n(n-1)/2}$. So, in total,
together with (\ref{Hermite}),
the conjugate of the l.h.s. of (\ref{gammaAT}) provides a factor
$(-1)^f$ with $f=n+1+(n+m-1)/2 + m(m-1)/2 + n(n-1)/2$. Noticing that
$f=(n+1) +(n-1)(n+1)/2 +m^2 /2=1 + n+  n^2 /2 + (m-1)(m+1)/2$, we find
$f$ is even (or odd) when $(m,n)=$(even, odd) (or (odd, even)). 
Therefore, the diagonal component $A^\pm$ (where odd forms appear) 
in  the r.h.s. of (\ref{gammaAT}) should be Hermitian, while $-i T$ and $-i\bar{T}$
(which are even forms) are related as $(-i T)^\dagger = +i \bar{T}$.} 
\begin{eqnarray}
 (A^{\pm})^\dagger =A^{\pm}, \,\,\,\,\, T^\dagger=\bar{T}
\end{eqnarray}
with the following definition of the conjugate which acts on the forms,
\begin{eqnarray}
 (dx^{\mu_1} \cdots dx^{\mu_n})^\dagger = (dx^{\mu_n} \cdots
dx^{\mu_1}).
\end{eqnarray}

%%%%%%%%%%%%%%%%%%%%%%%%%%%%%%%%%%%%%%%%

Let us write a more explicit form of the RR coupling formula.
As explained in \cite{Kraus:2000nj}, choosing an off-diagonal basis for gamma matrices,
the matrix multiplication rule is given by
\begin{eqnarray}
\begin{pmatrix}
A& B\\
C &D  
\end{pmatrix}
\begin{pmatrix}
A'& B'\\
C' &D' 
\end{pmatrix}
=
\begin{pmatrix}
AA'+(-)^{C'}BC'& AB'+(-)^{D'}BD'\\
(-)^{A'}CA'+DC' &(-)^{B'}CB'+DD'  
\end{pmatrix} ,
\end{eqnarray}
where $(-)^{A}$ is $+1$ or $-1$ if $A$ is bosonic or fermionic, respectively.
(Here $dx^\mu$ is treated as a fermion, but $\gamma^k$ is treated as a boson.)  
Thus we have
\begin{eqnarray}
2\pi i \, \mcf & =&  2 \pi i \,  (\rmd\mca-i \mca^2)
\nn\\
&=&
2\pi \begin{pmatrix}
-\bar{T}T+i(\rmd A^+ -iA^+A^+) & \rmd\bar{T}- i A^+ \bar{T}+ i\bar{T}A^-\\
\rmd T -i A^- T+iT A^+ & -T\bar{T}+ i(\rmd A^- - i A^-A^- )
\end{pmatrix}.
\end{eqnarray}
The RR coupling formula (\ref{generalRR}) is written with this curvature
$ {\cal F}$ of the superconnection, 
and here we provide 
an explicit form of $i \mcf$ as
\begin{eqnarray}
&&-\bar{T}T+i(\rmd A^+ -i A^+A^+) \nn\\
&&\quad
=-\bar{T}^{(0)}T^{(0)}\nn\\
&&\quad 
\quad+i\left( \rmd A^{(1)}_+ - i A^{(1)}_+\wedge A^{(1)}_+ \right)
 -  \bar{T}^{(0)}\wedge T^{(2)} -  \bar{T}^{(2)}\wedge T^{(0)} \nn\\
&&\quad 
\quad+i\left( \rmd A^{(3)}_+  - i A^{(1)}_+\wedge A^{(3)}_+ 
- i A^{(3)}_+\wedge A^{(1)}_+ \right) 
 -  \bar{T}^{(0)}\wedge T^{(4)} -  \bar{T}^{(4)}\wedge T^{(0)}
 -  \bar{T}^{(2)}\wedge T^{(2)}
\nn\\
&&\quad \,\,\,\,\,\,\,  +\cdots , \\
&&\rmd T - i A^- T+ i T A^+ \nn\\
&&\quad
 =\rmd T^{(0)}- i A_-^{(1)}\wedge T^{(0)}+ i T^{(0)}\wedge A_+^{(1)}\nn\\
&&\quad
\quad +\rmd T^{(2)} -i A_-^{(3)}\wedge T^{(0)} + i T^{(0)}\wedge A_+^{(3)}-i A_-^{(1)}\wedge T^{(2)}+ i T^{(2)}\wedge A_+^{(1)}\nn\\
&&\quad 
\quad+\cdots ,
\end{eqnarray}
where $T^{(n)}$ represents the $n$-form part.
Substitution of these expression explicitly 
with ${\rm Str} (*)={\rm Tr} (\sigma_3 (*) )$
provides the RR coupling. 

Instead of $\mca$, we can use $\tilde{\mca}$, which 
may be more physical.
For this, it would be more convenient to use following another definition
of the conjugate acting trivially on the forms:
$ (dx^{\mu_1} \cdots dx^{\mu_n})^\dagger = (dx^{\mu_1} \cdots
dx^{\mu_n})$.
With this and the non-trivial factor appeared in the transformation from
$\mca$ to $\tilde{\mca}$, 
we see that
\begin{eqnarray}
  (\tilde{A}^{\pm})^\dagger =\tilde{A}^{\pm}, \,\,\,\,\, \tilde{T}^\dagger=\bar{\tilde{T}}
\end{eqnarray}
for
\begin{eqnarray}
 \tilde\mca=\tilde\mca_{\mu_1,\cdots, \mu_n}^{k_1,\cdots, \mu_m} ({X})
dx^{\mu_1} \cdots dx^{\mu_n} \gamma^{k_1} \cdots \gamma^{k_m}
= \begin{pmatrix}
\tilde{A}^+({X}) & i^{\frac{3}{2}} \bar{T}({X})\\
i^{\frac{3}{2}} \tilde{T}({X}) & \tilde{A}^-({X})  
\end{pmatrix}.
\end{eqnarray}

The generalization to the D$p$-antiD$p$ brane system
can be done following \cite{Takayanagi:2000rz, AST} with the T-dualized
formula.
In this system the RR-charges are known to be generated by the 
ascent relation \cite{Asakawa:2001vm, Terashima:2001jc}
by the tachyon condensation and there would be analogue of 
this for the massive fields.
Our RR-charge formula is written using the supercommutator
which can be generalized to arbitrary numbers of D$p$-branes and  
anti D$p$-branes, in particular, which include the system 
with the BPS D$p$-branes only.
The validity of this procedure can be shown by
taking the some D-branes infinitely far away or
considering the infinitely many D-anti-D-brane system
realized by the Gamma matrix of $SO(\infty)$ 
with the tachyon condensation which gives any number of D-branes.
Note that, of course, 
restricting the super connection
${\cal A}$ to its upper-left corner, we trivially reproduce the 
(the non-Abelian generalization of) 
RR coupling formula (\ref{CSterm2}) of the BPS D-brane.

%%%%%%%%%%%%%%%%%%%%%%%%%%%%%%%%%%%%%%%%

\section{Summary and discussion}

In this paper, we found the most general RR coupling formula of D-branes.
We allowed arbitrary number of insertions of all massive excitations of
open superstring theory. The worldsheet theory is the 2-dimensional 
${\cal N}=(1,1)$ supersymmetric field theory with free chiral multiplets. The
worldsheet was taken to be a cylinder, and at one boundary we considered
arbitrary boundary deformations preserving a half of the supersymmetries,
and the other boundary corresponds to the Ramond-Ramond bulk vertex.
The localization technique is powerful enough to evaluate the cylinder
partition function with these boundary conditions, and it is nothing but the
RR coupling formula in view of boundary superstring field theory (BSFT).

We considered BPS D9-branes and also the case of pairs of a D9 and an antiD9. 
In either case, 
the RR charge 
formula is written in a simple manner with Quillen's superconnection, 
(\ref{generalRR}). 
Interestingly, only a finite number of massive excitation fields can enter the
formula, and the resultant Chern characters of the superconnection can make
sure the RR charge quantization.

We have used a cylinder worldsheet, instead of a disk worldsheet which 
has been commonly used in BSFT. 
Instead, we used the infinitely long 
cylinder which is an alternative flat worldsheet,
to formulate the off-shell supersymmetries consistent with the 
standard worldsheet theory.
In any case, the shape of the worldsheet does not give physical difference 
for the result, since normally the deformation of the shape should correspond
to a field redefinition on the worldsheet boundary fields.

In this paper we concentrated on the RR coupling formula. 
From the viewpoint of the BSFT, equally important action is for the Neveu-Schwarz Neveu-Schwarz (NSNS) sector
which describes couplings of 
all the off-shell open superstring modes to the bulk gravity.
To describe the NSNS sector, one needs a different quantization of fermion fields
on the worldsheet.

We showed also that it is impossible to create of a D9-brane by a condensation of massive modes
on a BPS D9-brane or on a D9-antiD9 pair. Our observation is in a good
contrast with a recent report \cite{Erler:2014eqa}
on multiple D-brane solutions in cubic open string field theory \cite{Witten:1985cc}. Since no good relation (field redefinition) between the BSFT and 
the cubic open string field theory, it is plausible that the theory
configuration space allowed for each theory is different from the first
place.\footnote{
We can think a BPS D-brane as a configuration in 
the infinitely many D9-antiD9-brane pairs with the nonzero tachyon.
Then, in this point of view,
a creation of the any number of D9-branes can be possible 
in the BSFT.}
It would be interesting to find out where the discrepancy comes from, and
a more precise understanding of the 
off-shell configuration space of superstring theory.

Our most general Ramond-Ramond coupling formula is written by higher form fields.
It is interesting that the higher Chern characters of massless gauge field strengths
are combined with the higher form fields on an equal footing. Our work showed
for the first time 
that the open string higher forms, which are massive excitations, are relevant to
extended objects in string theory --- D-branes. In view of recent progress on generic study of higher form fields \cite{Gaiotto:2014kfa}, it is important to study the higher structure of open string theory further.

%%%%%%%%%%%%%%%%%%%%%%%%%%%%%%%%%%%%%%%%

\section*{Acknowledgments}

S.~S. would like to thank Satoshi Yamaguchi for discussions.
S.~T. would like to thank Shigeki Sugimoto for a useful comment.
The work of K.~H.~is partially supported by the RIKEN iTHES project.
The work of S.~S.~is supported by the JSPS Fellowship for Young Scientists.
The work of S.~T.~is partly supported by JSPS KAKENHI Grant Number 23740189.

%%%%%%%%%%%%%%%%%%%%%%%%%%%%%

\end{document}